\newcommand{\refsec}[1]{{Section~\ref{#1}}}
\newcommand{\reffig}[1]{{Figure~\ref{#1}}}
\newcommand{\reftab}[1]{{Table~\ref{#1}}}
\newcommand{\refeqn}[1]{{Equation~(\ref{#1})}}

\long\def\comment#1{}
\long\def\ignore#1{}
\long\def\note#1{}

\documentclass{sig-alternate}

\begin{document}
%
\conferenceinfo{SIGKDD}{'12}
\title{Dynamic Multi-Relational Chinese Restaurant Process \\
for Analyzing Influences on Users in Social Media}

%
%
%
%
%

\numberofauthors{3} 
%
\author{
%
%
\alignauthor
Himabindu Lakkaraju \\
       \affaddr{IBM Research - India}\\
       \email{katynaga@in.ibm.com}
\alignauthor
Indrajit Bhattacharya\\
\affaddr{Indian Institute of Science}\\
\email{indrajit@csa.iisc.ernet.in}
\alignauthor
Chiranjib Bhattacharyya \\
\affaddr{Indian Institute of Science}\\
\email{chiru@csa.iisc.ernet.in}
}
\maketitle
\begin{abstract}
We study the problem of analyzing influence of various factors affecting individual messages posted in social media.
The problem is challenging because of various types of influences propagating through the social media network that act simultaneously on any user. 
Additionally, the topic composition of the influencing factors and the susceptibility of users to these influences evolve over time.
This problem has not studied before, and off-the-shelf models are unsuitable for this purpose.
To capture the complex interplay of these various factors, we propose a new non-parametric model called the Dynamic Multi-Relational Chinese Restaurant Process.
This accounts for the user network for data generation and also allows the parameters to evolve over time.
Designing inference algorithms for this model suited for large scale social-media data is another challenge. 
To this end, we propose a scalable and multi-threaded inference algorithm based on online Gibbs Sampling.
Extensive evaluations on large-scale Twitter and Facebook data show that the extracted topics when applied to authorship and commenting prediction outperform state-of-the-art baselines.
More importantly, our model produces valuable insights on topic trends and user personality trends, beyond
the capability of existing approaches.
\end{abstract}


\section{Introduction}\label{intro}

Social networking sites, such as Twitter, Facebook, MySpace etc, have proven to be extremely popular platforms for users for sharing views and opinions using short posts\footnote{Our work focuses on such micro-blogging sites.}. 
Understanding and analyzing topics in social media has become immensely important for a variety of stakeholders, such as companies advertising products and identifying customer segments, social scientists and national security agencies, leading to a surge in research interest \cite{recsys,chi,www,wen10,amrkdd11,dramage10,jure}.
There are two major distinguishing features of social media data.
First, users are influenced by a variety of factors when posting messages. 
The four major factors have been identified to be personal preferences of the users, their immediate network of friends on the network, geographic or regional issues and events and world-wide happenings \cite{jure}.
While all factors typically affect all users, different users have different `personalities', in that they are influenced by these factors in different degrees.
Secondly, social media data is inherently dynamic. 
Topics follow different `trends'; individual interests of influential users, or issues starting off within a small network of friends, sometimes lead to global upheavals, while other enjoying global popularity are slowly relegated to individual favorites.
Similarly, user personalities also evolve and show trends over a geography or sub-network.

Owing to these multitude of factors, and the intrinsic interplay between them, analysis of social media data has been a major challenge.
Most existing approaches fall short of addressing the problem in its entirety, and only model isolated factors and their interactions \cite{dramage10,wen10,amrkdd11}.
A major hurdle for sophisticated models is the scale of the data; the associated inference algorithms need to be scalable and efficient.

In this paper, we propose a non-parametric probabilistic approach for analyzing social media data.
Specifically, we first propose an augmentation of the Chinese Restaurant Process\cite{pitman02}, called the Multi-Relational Chinese Restaurant Process (MRelCRP), that accommodates users and multiple relationships over them, for assigning topics to posts.
By using relationships, the MRelCRP defines a new and different family of distributions compared to the traditional non-parametric processes such as the Dirichlet Process \cite{antoniak:aos74}, and its hierarchical versions \cite{teh06}.
We further propose a dynamic version of the MRelCRP (D-MRelCRP) that allows temporal evolution of the model parameters to capture topic and personality trends.
The rich interactions of various parameters in the model are able to capture the various interplays in social media data.
Crucially, we propose an efficient and multi-threaded algorithm, based on online collapsed Gibbs sampling, for performing learning and inference for Dynamic MRelCRPs. 

We evaluate the proposed model on two large scale datasets. The first dataset consists of 360 million posts from Twitter. The second dataset consists of 300K
posts from Facebook.
We demonstrate both qualitatively and quantitatively the goodness of the topics discovered by our model.
When employed for predicting authorship and user activity, models using these topics significantly outperform state-of-the-art baselines.
More importantly, our model is able to discover interesting and insightful topic and personality trends.
For example, our analysis shows that users posts are mostly influenced by personal preferences, rather than global, regional or social-network factors, except in times of major world events, when users become swayed by global influences at the cost of personal preferences. 
We are not aware of any existing model that can perform such a wide array of analyses effectively on social media data.


The rest of the paper is structured as follows. 
We discuss related work on social media analysis and topic models in \refsec{rw}.
We describe our proposed model in \refsec{model} and the associated inference algorithm in \refsec{inf}.
Experimental results are presented in \refsec{exp} and we conclude in \refsec{conc}.

\section{Related Work}\label{rw}
Here, we discuss our contributions in the light of related work in non-parametric probabilistic modeling and social media analysis.

{\bf Non-parametric models: }
The Dirichlet Process (DP) \cite{antoniak:aos74} is a prior over a countably infinite set of atoms, and is popularly used as a prior for mixture models (DP Mixture Model) in applications, where the number of clusters is difficult to provide as a parameter. 
The Chinese Restaurant Process \cite{pitman02} provides a generative description for the Dirichlet Process, and is useful for designing sampling algorithms for DP mixture models.
The distributions defined by these models are exchangeable, in that different permutations of the data are equally probable.

CRPs have been extended to handle distances and relations. 
The distance dependent CRP (DD-CRP) \cite{blei:icml10} takes into account a distance matrix over the input data points. 
Unlike the DP-HDP family, this results in a distribution that is not exchangeable, which is a feature of many applications.
In comparison, the RelCRP uses an additional non-unique label for each data point, and a general graph defined over them.
The resultant distribution is exchangeable.
As such, the DD-CRP and the RelCRP define different families of distributions, and one cannot be represented by the other.
There is different body of work that use CRPs for modeling relations \cite{kemp:aaai06} and their dynamic evolution \cite{ishiguro:nips10}. 
These are unsuitable for our current application, where we do topical analysis of the data points, based on relations between their (user) labels.

Many applications require multiple coupled Dirichlet Processes. 
The Hierarchical Dirichlet Process (HDP) \cite{teh06} is one way to introduce coupling using a two level structure.
The HDP can be useful, for example, for extending the popular Latent Dirichlet Allocation (LDA) model \cite{blei03}, for countably infinite number of topics \cite{amrkdd11}.
The HDP can be equivalently represented by an extension of the CRP called the Chinese Restaurant Franchise (CRF) \cite{teh06}. 
Just as the CRF introduces coupling between CRPs, the MultiRelCRP introduces coupling between RelCRPs. 
However, the nature of the coupling in the MultiRelCRP can be much richer, depending on the relationships, as we explain in \refsec{subsec:mrelcrp}.  

Temporal evolution has been addressed in the context of non-parametric models \cite{ahmed:sdm08,ahmed:2010,amrkdd11}, where the parameters of the the corresponding static model become functions of time. 
Some of the approaches are amenable to scalable inference, while others are not.
For the Dynamic MRelCRP, we use the dynamic evolution of the parameters proposed in the context of Recurrent CRF \cite{ahmed:2010,amrkdd11}, because of the scalability of the associated inference problem.
Note, however, that the similarity between the Recurrent CRF and the Dynamic-MRelCRP is only in the temporal evolution of model parameters.
The static model is a HDP/CRF, as compared to the RelCRP in our case.

{\bf Social Media Analysis: }
There has been a surge of literature on problems involving social media content. 
Work has mostly been focused around 
(a) Content analysis on microblogs, (b) Inferring user interests and (c) Mining patterns of variation on social media, as we discuss below.

(a) 
Most content analysis papers \cite{soma10} use standard topic models such as LDA \cite{blei03} or basic metrics like tf-idf. 
Focusing on the specific content of miroblogs, Ramage et. al. \cite{dramage10}  proposed an LDA variant that accounts for hashtags in content analysis. 
One problem with this approach is that hastags are not general features of social media data, and are often unreliable.
There is little modeling work that takes into account the rich features of social media such as network, geography, etc.

(b) 
In the context of microblogging sites, content recommendation approaches \cite{recsys,chi,www} assessing user interests based on their activities. 
Recently, Wen et. al \cite{wen10} have proposed an approach which studies the influence of the network on users. Ahmed et. al. \cite{amrkdd11} model the dynamics of user interest and also the account generic popularity of a particular item, but do not consider the influence of various external factors like network of users and geography. 
Thus most of related work either deals with the influence of a single factor or user preferences. 

(c) 
Yang et. al. \cite{jure} made one of the first attempts at understanding the temporal evolution of patterns on social networking sites like Twitter. Apart from temporal dynamics, study of such patterns with respect to geography and other factors has not been explored for content on social networking sites. 


\section{Model for Social Media}\label{model}
In this section, we describe our Dynamic Multi-Relational Chinese Restaurant Process model, which we employ to study the interplay of world-wide, geographic, network and user specific factors, and their dynamics, in social 
media.
We build up our model in steps, first describing the static Relational Chinese Restaurant Process, then incorporating multiple relations, and finally adding temporal evolution to it.
In our application, the basic task is to associate topics with individual posts or tweets.
The topics correspond to concepts such as `movies', `sports', `politics' etc.
Unlike topic models such as LDA \cite{blei03}, which associate a distribution over topics with each document,
we assume that each post, considering its shortness, corresponds to exactly one topic.
This makes the model simpler and the associated inference algorithm more efficient and scalable.

\subsection{Relational Chinese Restaurant Process}

The Dirichlet Process \cite{antoniak:aos74} has become a popular non-parametric prior in clustering applications, where the number of clusters is not needed to be specified apriori, but instead grows with the data size. 
The Chinese Restaurant Process (CRP) \cite{pitman02} provides a fanciful description of the Dirichlet Process, by imagining data points as customers being seated at tables, which represent clusters, as they enter the restaurant. 
Let $w_i$ denote the $i^{th}$ data point, or post in our case, and $z^i$ denote the cluster (or table assignment) for the post. 
Then, given the assignments $z_{1:i-1}$ of the first $i-1$ customers to $K$ tables, the conditional distribution for the table assignment of the $i^{th}$ customer is given by the CRP as follows:
\begin{eqnarray}\label{eqn:crp}
P(z_{i} = k | z_{1:(i-1)}, \alpha) & \propto n_{k} & k\leq K \nonumber \\
 & \alpha & k= K+1 
\end{eqnarray}
where $n_k$ is the number of customers already assigned to table $k$.
Note that this a `rich gets richer' model, where tables with more customers have a higher probability of getting new customers, but new tables also have a non-zero probability ($\alpha$) of getting customers.

When each table $i$ is associated with a (topic) distribution, with parameters $\phi_i$ drawn iid from an appropriate base distribution $H$, the CRP can be used as a prior for mixture distributions. 
Once the $i^{th}$ customer is seated at a table $z_i$, the corresponding data item $w_i$ can be drawn independently from the distribution $\phi_{z_i}$ associated with the table. 
For a generative model for posts, each topic distribution can be a multinomial $Mult(\phi_i)$ over the post vocabulary, so that each word $w_{ij}$ of the post is generated independently from that topic, and the base distribution $H$ can be chosen to be a Dirichlet $Dir(\beta)$, since it is conjugate to the multinomial.

Though defined as a sequential process, the CRP mixture model can be easily shown to be exchangeable, which means that all permutations of observed data $\{w_i \}$ have the same probability under the model.
The Chinese Restaurant Process has been widely used in generative models for different applications \cite{teh06,sudderth:nips05,xing:jcb07}. 
However, it is unsuitable for social media data, since it ignores a fundamental aspect --- the social network over users who generate the content. 
Specifically, each post has associated with it a user variable $u_i$, that takes values from a finite set of users $\cal U$. 
These users are further connected by a network of relationships.
To accommodate this, we augment the Chinese Restaurant Process to handle such relationships. 

In the Relational Chinese Restaurant Process (RelCRP), each customer (data point) is associated with a label $u_i\in \cal U$. 
In the context of social media data, we will refer to each element in $\cal U$ as a user, and say that each data point or post has a user label.
In addition, we have a relationship $\cal R$, such that each element $r\in \cal R$ is a subset of $\cal U$. 
We can imagine $\cal R$ as defining a set of hyper-edges over elements in $\cal U$.
Note that we do not fix the candinality of the elements in $R$.
We will see the need for this shortly.
Using $\cal R$, we can define the neighbors $N(u,R)$ of an element $u\in \cal U$ as all other elements that share a relation with $u$: $N(u,R) = \{u'\in {\cal U}: \exists r\in {\cal R}, u\in r, u'\in r\}$.

Given the additional $u_i$ labels and the relationship $\cal R$, the conditional distribution of the table assigned to the $i^{th}$ customer is defined in RelCRP as follows:
\begin{eqnarray}\label{eqn:rcrp}
P(z_{i} = k | z_{1:(i-1)}, u_{1:i}, \cal R, \alpha) & \propto n_{k}^{N(R,u_i)} & k\leq K \nonumber \\
 &  \alpha & k= K+1 
\end{eqnarray}
where $n^{N(R,u_i)}_k$ is the number of neighbors of $u_i$ in $\cal R$ already assigned to table $k$.

Let us now look at some example uses of the RelCRP in the context of social media data. 
We start from the trivial case, where the RelCRP reduces to the CRP, and then move on to more interesting ones.

{\bf Influence of World-wide Factors: }
Very commonly users are influenced by globally popular events or entities when choosing a post topic.
For example, users who are not fans of Michael Jackson tweeted on this topic in the event of his unexpected death.
This can be captured in the RelCRP by associating a unique label $u_i$ with each data point, along with a `complete' relationship ${\cal R}_w$, that contains a single relation (hyper-edge) over all $u\in {\cal U}$.  
In this case, \refeqn{eqn:rcrp} reduces to:
\begin{eqnarray}\label{eqn:rcrp-g}
P_w(z_{i} = k | z_{1:(i-1)}, u_{1:i}, {\cal R}_w, \alpha) & \propto n_k & k\leq K \nonumber \\
 & \alpha & k= K+1 
\end{eqnarray}
where $n^{N({\cal R}_w,u_i)}_k = n_k$ is the number posts by {\it all users} (which is the neighbor set of $u_i$) already assigned to table (topic) $k$. 
Note that this is the same as \refeqn{eqn:crp}. 
Thus the RelCRP is able to recover the traditional CRP, using unqiue data labels and a `complete relationship'.

{\bf User's Personal Preferences: } 
One of the most significant factors influencing the content of a post is the preference of the associated user.
A specific user may be more interested in `movies' that in `sports' or `politics'.
Evidence of this can be found in the topics of this user's earlier posts --- a user is more likely to post on a topic that she has used more frequently.
To capture this in the RelCRP, we set $u_i$ to be user identifier, and simply construct an empty relation ${\cal R}_u$ over $\cal U$. 
Given $({\cal U}, {\cal R}_u)$, \refeqn{eqn:rcrp} reduces to:
\begin{eqnarray}\label{eqn:rcrp-u}
P_u(z_{i} = k | z_{1:(i-1), u_{1:i}}, {\cal R}_u, \alpha) & \propto n^{u_i}_k & k\leq K \nonumber \\
 & \alpha & k= K+1 
\end{eqnarray}
where $n^{N({\cal R}_u,u_i)}_k = n^{u_i}_k$ is the number of posts by user $u_i$ (who is her only neighbor) already been assigned to table $k$.
Note that even the case of the empty relation ${\cal R}_u$ cannot be captured by the traditional CRP.

{\bf Influence of Friend Network: }
A user is often influenced by the post topics of her friends.
To capture this, as before, we set the label $u_i$ of the post to be the user id, and construct ${\cal R}_n$ based on the friendship network: for each follower or friendship relation between users $u_i$ and $u_j$, we add a tuple $(u_i,u_j)$ to ${\cal R}_n$.
Note that in this case all elements of ${\cal R}_n$ have cardinality $2$.
Given $({\cal U},{\cal R}_n)$, $P_n(z_{i} = k | z_{1:(i-1)}, u_{1:i}, {\cal R}_n, \alpha)$ has the same form as \refeqn{eqn:rcrp}, where $n^{N({\cal R}_n,u_i)}_k$ is now the number of times followees or friends of user $u_i$ have posted on topic $k$.

{\bf Influence of Geography: }
As a final example, a user's posts may also be influenced by geographic trends. 
For instance, an national election draws a lot of attention from citizens of that country. 
This can be captured by the RelCRP, by again associating labels $u_i$ with user id's, and constructing ${\cal R}_g$ to capture geographic locations: adding a hyper-edge in ${\cal R}_g$ over all users in a specific country. 
Typically, the geographic location can be known from the profile of the user, and we assume such a construction to be possible. 
Note that in this case every edge has a different cardinality, and
most will be extremely large. 
Interestingly, the RelCRP does not require maintaining the explicit relations, but only simple statistics over them.
Given $({\cal U},{\cal R}_g)$, $P_g(z_{i} = k | z_{1:(i-1)}, u_{1:i}, {\cal R}_g, \alpha)$ again takes the form of \refeqn{eqn:rcrp}, where $n^{N({\cal R}_g,u_i)}_k$ is now the number of times users in the same geography as user $u_i$ have posted on topic $k$.

Thus, the RelCRP can be used to capture the different posting patterns in social media within a single framework, in a way that the traditional CRP cannot.
Just like the CRP, however, the RelCRP can be used to define a mixture model by associating a topic with each table. 
It can be shown that the resultant distribution remains exchangeable.

\subsection{Multi-Relational CRP}\label{subsec:mrelcrp}
We have seen that the RelCRP is able to model the individual effect of the world-wide factors, user preferences, friend network and geographic factors when the labels and relationships are appropriately defined. 
However, in reality, all of these influences act {\it simultaneously} on any user, and their interplay determines the content of each of her posts.
Further, this aggregate influence pattern is user-specific. 
For example, different users are affected differently by the same combination of world and geographic events.
We now present the Multi-Relational Chinese Restaurant Process (MRelCRP) that captures such aggregate influences using multiple relations defined over the same user labels. 

The MRelCRP is characterized by a set of labels $\cal U$, along with $m$ relations $\{{\cal R}_i \}_{i=1}^m$ defined over $\cal U$.
With the $i^{th}$ data point (post), we associate an additional variable $f_i$, which takes values from $\{1\ldots m\}$, indicating the relationship that influenced this data point.
This depends on the associated label (user) $u_i$. 
For each label $u\in {\cal U}$, there is a $m$-dimensional multinomial distribution $Mult(\pi_u)$. 
Each $\pi_u$ is assumed to be generated iid from a Dirichlet $Dir(\alpha)$. 
We interpret $\pi_{uj}$ as the probability of label $u$ being influenced by the $j^{th}$ relationship ${\cal R}_j$, i.e. $P(f_i=j|u_i=u)$. 
We may imagine $\pi_u$ as reflecting the `personality' of user $u$.
Given these parameters, and the assignment of the first $i-1$ posts to $K$ topics, the MRelCRP defines the conditional distribution of the topic assignment of the $i^{th}$ post with label $u_i$ as follows:
\begin{eqnarray}\label{eqn:mrcrp}
& & P(z_i=k | z_{i:i-1}, u_{1:i}, \alpha, \{{\cal R}_j\}, \{\pi_u\}) \nonumber \\
& = & \sum_{j} \pi_{u_ij} P(z_i=k | z_{i:i-1}, u_{1:i}, \alpha, {\cal R}_j)
\end{eqnarray}  
which is a mixture of $m$ individual RelCRP distributions, defined according to \refeqn{eqn:rcrp}.
This can be interpreted as first selecting a particular RelCRP from a prior distribution {\it specific to the label} $u_i$, and then selecting a table using the selected RelCRP.

The aggregated influences in the post generation process can now be captured by the MRelCRP framework, by considering the set of $4$ relationships $\{{\cal R}_w, {\cal R}_u, {\cal R}_n, {\cal R}_g \}$.
A $4$-dimensional influence factor $\pi_u$ is sampled for each user $u$ from $Dir(\alpha_w, \alpha_u, \alpha_n, \alpha_g)$.
This can be imagined to reflect the personality of the user. 
Then, for each post, a topic is selected for it, in two steps, using \refeqn{eqn:mrcrp}.
Finally, the individual words in the post are sampled iid from this selected topic. 
This is described in \reftab{tab:gmstatic}.

\begin{table}[h]\label{tab:gmstatic}
\caption{Generative Process for MRelCRP}\label{tab:mfcrp}
\begin{center}
\begin{tabular}{|l|} \hline
1. For each topic $k$ \\
\hspace{.15in} a. Sample $\phi_k \sim Dir(\beta)$ \\
2. For each user $u$ \\
\hspace{.15in} a. Choose $\pi_u \sim Dir(\alpha_w, \alpha_u, \alpha_n, \alpha_g)$ \\
3. For each post $i$  \\
\hspace{.15in} a. Choose $f_{i} \sim Mult(\pi_{u_i})$\\ 
\hspace{.15in} b. Choose $z_{i} \sim P(z_i|z_{1:i-1},u_{1:i},\alpha,{\cal R}_{f_i})$ \\ 
\hspace{.15in} c. For the each word $j$ of post $i$ \\
\hspace{.25in} i. Choose $w_{ij} \sim Mult(\phi_{z_{i}})$ \\ \hline
\end{tabular}
\end{center}
\end{table}

{\bf Couplings in the MRelCRP: }
It is important to observe the coupling that the MRelCRP creates between different RelCRPs, that helps capture the interplay of various factors for social media.
(a) First, we analyze the dependencies for a single relationship ${\cal R}_i$.
Observe that there are $N$ RelCRPs, one for each user (label). 
However, all of these $N$ RelCRPs need not be distinct.
This depends on the nature of the relationship.
For example, in the setting above, ${\cal R}_w$ is a `complete' relationship. 
As a consequence, the neighbor sets are the same for all users, and the world-wide RelCRP is identical for all users.
For the geographic relationship ${\cal R}_g$, since the individual relations do not overlap, the geographic RelCRP is identical for all users from the same country.
This creates one type of dependence across users.
In contrast, for the friend relationship ${\cal R}_n$, in general distinct users have different sets of friends, and their RelCRPs are distinct. 
However, they are still coupled, since the underlying topics are the same, and a post by user $u$ on topic $k$ increases the count $n^{N({\cal R}_n,u')_k}$ for all friends $u'$ of $u$.
Thus, for all of these three relationships, evidence can flow {\it between users} over hyper-edges in the relationship.
Finally, for the user preference relationship ${\cal R}_u$, the RelCRP for each user is distinct, and there are no dependencies.
(b) Now, we analyze the new dependencies that are created when multiple relationships are coupled in the MRelCRP. 
Observe that for $m$ relationships, there is a total of $m\times N$ RelCRPs, $m$ for each user, but all of these need not be distinct, as above.
The $m$ distinct RelCRPs for each user are now coupled; evidence can flow {\it between relationships} through the users.
In the context of social media, this leads to interplay between world-wide, geographic, network factors and personal preferences.


\subsection{Dynamic Multi-relational CRP}

The two key distinguishing aspects of social media data are the network structure, and the dynamic nature of the topics and user influence patterns or personalities.
The MRelCRP captures the network aspect, but falls short on the second count.
Before extending our model, we first enumerate the different aspects of the data that evolve with time.
(a) The number of topics changes as old topics die out and new topics are born.
(b) Popularity of topics change, world-wide, in specific geographies, sub-networks and in the preferences of individual users. We call these topic trends.
(c) User personalities change, and they become more or less susceptible to being influenced by world-wide, geographic, network and individual preferences.
(d) Existing topics also evolve as new words enter the vocabulary and existing words go out of fashion.
We now propose the Dynamic Multi-relational Chinese Restaurant Process (D-MRelCRP) that accounts for all of these temporal evolutions.
In reality, the number of users also change over time and the network grows or shrinks, but we do not consider this aspect in our current model.

We assume that the data has been segmented into epochs, or in other words, each data element is labeled with a time-stamp that takes values from $\{1\ldots T\}$.
In practice, epochs may be appropriately defined (eg. hour, day, week, etc) depending on the application.
The Dynamic MRelCRP consists of one MRelCRP for each epoch.
We introduce dependencies between the parameters of the MRelCRPs across epochs to capture the different aspects of temporal evolution, as we describe next. 
We use additional subscripts on parameters to indicate epochs. 

Note that individual RelCRP's naturally allow the number of topics to change. 
We do not need to address this separately in the D-MRelCRP.

\subsubsection{Topic Trends}
Different topics have different trends, in that some start out being popular in certain geographies, to being global hits.
Some others may start as preferences of influential individual users and evolve to regional or world favorites.
To capture this, topic popularities in our model need to change over epochs. 
Since we have modeled popularity of topics using counts, to make this approach dynamic, topic counts of specific epochs are made dependent on those of earlier epochs, following the approach of \cite{amrkdd11}. 
We extend the basic RelCRP conditional distribution (\refeqn{eqn:rcrp}) with epoch indices as follows:
\begin{eqnarray*}
P_t(z_{i} = k | z_{1:(i-1)}, u_{1:i}, {\cal R}, \alpha) \propto & n_{k,t}^{N(R,u_{i})} + \bar{n}_{k,t}^{N(R,u_{i})}\ \ \  k\leq K \\
 & \alpha \ \ \ k= K+1 
\end{eqnarray*}
where $n^{N(R,u)}_{k,t}$ is the number of neighbors of $u$ in $\cal R$ already assigned to table $k$ in the $t^{th}$ epoch, while $\bar{n}^{N(R,u)}_{k,t}$ captures the historical counts in recent epochs, with exponentially decaying weights, as follows:
\begin{eqnarray*}
\bar{n}^{N(R,u)}_{k,t} = \sum_{\delta=1}^{\Delta} e^{-\delta/\lambda} n^{N(R,u)}_{k,t-\delta} 
\end{eqnarray*}
where $\lambda$ is the decay factor.
The MRelCRP for $t^{th}$ epoch is now defined using a mixture of such RelCRP conditionals as in \refeqn{eqn:mrcrp}.

\subsubsection{User Personality Trends}
It is natural for user personalities to be time dependent as well.
A user may become more susceptible to the influence of her friends and deviate from her earlier personal preferences.
In the MRelCRP framework, this corresponds to the mixture distribution $\pi_u$ for each user $u$ being a function of the epoch.
Recall that each $\pi_u$ is sampled iid from a Dirichlet prior $Dir(\alpha_w,\alpha_u,\alpha_n,\alpha_g)$. 
We introduce a temporal dependence by adding a dynamic component to the prior parameter, in the spirit of \cite{amrkdd11}, as follows: 
\begin{eqnarray*}
& \pi_{u,t} \sim Dir(\alpha_w+\bar{\alpha}_{u,w,t},\alpha_u+\bar{\alpha}_{u,u,t},\alpha_n+\bar{\alpha}_{u,n,t},\alpha_g+\bar{\alpha}_{u,g,t}) \\
& \bar{\alpha}_{u,f,t} = \sum_{\delta=1}^{\Delta} e^{-\delta/\lambda} m_{u,f,t-\delta}
\end{eqnarray*}
for $f\in \{w,u,n,g\}$, and $m_{u,f,t}$ being the number of times user $u$ was influenced by relationship $f$ in epoch $t$.

\subsubsection{Evolving Topic Distributions}
The topic-word distributions $\phi_j$ also evolve with time. 
To capture this, we again introduce a temporal dependence in the prior distribution.
Specifically, each topic distribution $\phi_{k,t}$ is now sampled from $Dir(\beta_{k,t} + \beta)$.
The element $\beta_{k,w,t}$ of dynamic component $\beta_{k,t}$ depends on how frequently the word $w$ in the vocabulary has been historically observed under topic $k$ until epoch $t-1$:
\[
\bar{\beta}_{k,w,t} = \sum_{\delta=1}^{\Delta} e^{-\delta/\lambda} m_{k,w,t-\delta} 
\]
where $m_{k, w, t}$ corresponds to the number of times word $w$ is associated with the topic $k$ in epoch $t$.

These three dynamic dependences introduced between the parameters of the MRelCRPs corresponding to different epochs, defines our complete D-MRelCRP.
\comment{
 }

\section{Inference}\label{inf}
In this section, we discuss the key challenges in performing inference for the proposed D-MRelCRP model, and present our inference algorithm addressing these challenges. 
The inference problem involves determining the posterior distribution over the two latent variables variables, the topic label $z_{i,t}$ and the influence variable $f_{i,t}$, for all posts $i$ in all epochs $t$. 
The parameter estimation problem involves finding the posterior distribution of the model parameters, the topic distributions $\phi_k$ and the personalities $\pi_u$ of the users.
The two problems are coupled, and solving them exactly is intractable \cite{blei03}. 
We resort to approximate techniques based on collapsed Gibbs sampling. 
However, the traditional approach \cite{canini}, where the topic and influence labels of {\it each post} are {\it repeatedly sampled} until convergence from the conditional distributions given all other labels, is infeasible for us given the size of the data.
Even Sequential Monte Carlo methods \cite{canini}, that rejuvenate a few older labels, are infeasible.
We adopt the online algorithm \cite{banerjee}, which was proposed for parametric models, and modify it appropriately for our model.
In this approach, earlier labels are not revisited. 
This allows the algorithm to scale, at the expense of sub-optimal estimates at the beginning, and is also concordant with the online nature of social media data \cite{amrkdd11}.
Before describing the details, we first describe the conditional distributions that are required by the algorithm.


{\bf Conditional Distributions: }
In the online setting, the distribution for the influence factor $f_i$ for the $i^{th}$ post is conditioned on the topic and influence labels of all earlier posts, their user labels and the content of the current post. 
For the Dynamic MRelCRP, this looks as follows:
\begin{eqnarray}\label{eqn:dmrcrp-f}
P(f_i = f| \alpha, u_i, z_{1:i}, f_{1:(i-1)}, \textbf{w}_{1:i}, {\cal R}) \nonumber \\
\propto (m_{u,f,t}+\bar{\alpha}_{u,f,t}+\alpha_f) \times (n_{k,t}^{N({\cal R}_f,u_i)} + \bar{n}_{k,t}^{N({\cal R}_f,u_i)}) 
\end{eqnarray}
where, $\alpha\equiv \{\alpha_w, \alpha_u, \alpha_n, \alpha_g\}$, and the counts are as defined in \refsec{model}.

The conditional distribution for topic label $z_i$, additionally conditioned on influence factor $f_i$, is given by:
\begin{eqnarray}\label{eqn:dmrcrp-z} 
& P(z_i = k | f_i, \beta, z_{1:(i-1)}, u_{1:i}, \textbf{w}_{1:i}, {\cal R}, \alpha ) \nonumber \\
& \propto  (n_{k,t}^{N({\cal R}_{f_i},u_i)} + \bar{n}_{k,t}^{N({\cal R}_{f_i},u_i)}) \prod_{l=1}^{N_i} \frac{n_{k,v,t} + \bar{\beta}_{k,v,t} + \beta}{\sum_{r=1}^V (n_{k,r,t} + \bar{\beta}_{k,r,t} + \beta)} \ \   k \leq K \nonumber \\ 
 & \propto  \alpha \prod_{l=1}^{N_i} \frac{n_{k,v,t} + \bar{\beta}_{k,v,t} + \beta}{\sum_{r=1}^V (n_{k,r,t} + \bar{\beta}_{k,r,t} + \beta)}  \ \  k = K+1  
\end{eqnarray}
where, $w_{il}$ corresponds to the $v^{th}$ word of the vocabulary, $n_{k,v,t}$ corresponds to the number of times $v^{th}$ word in the vocabulary is associated with topic $k$ during epoch $t$. 
Note that online inference, the counts in the equations above correspond to the data instances (posts) which have arrived before the $i^{th}$ instance.
Also, the conditional distributions for the static model (MRelCRP) can be obtained by removing the historical counts in the above expression, specifically by setting $\bar{\alpha}_{u,f,t}=\bar{n}_{k,t}^{N(R,u_i)}=0$ in \refeqn{eqn:dmrcrp-f}, and $\bar{n}_{k,t}^{N(R,u_i)}=\bar{\beta}_{k,v,t}=0$ in \refeqn{eqn:dmrcrp-z}.
Similarly, the conditional distribution for RelCRP, which has a single relationship $\cal R$ can be obtained as a special case of the MRelCRP, by taking counts $n_{k,t}^{N(R,u_i)}$ with respect to $\cal R$.

{\bf Parallel Inference Algorithm: }
A straight-forward online algorithm, that makes a single sequential pass over the data, is infeasible considering the scale of social media data.
This necessitates a parallelized inference algorithm. 
Sampling based parallel inference algorithms for hierarchical bayesian models has received attention in the literature \cite{maxweling,alex}.
These approaches split data across threads or processors, execute Gibbs iteration on them independently, and finally, consolidates labels across threads asynchronously at the end of each iteration.
In contrast, parallelization of our algorithm results in {\it independent, online updates} in each thread.
Additionally, D-MRelCRP being a non-parametric model, new topics are created by each thread, and in the absence of repeated Gibbs iterations, are not sufficiently consolidated.
As a result, we require a synchronous architecture, where all new topics are explicitly consolidated by a master thread at the end of each iteration.  
Our multi-threaded inference algorithm is described in \reftab{tab:algo}.

\begin{table}[h]
{\scriptsize
\caption{Parallel Inference Algorithm}\label{tab:algo}
\begin{center}
\begin{tabular}{|l|} 
\hline
Master Thread \\
1. \hspace{.1in} 	Read first N posts \\
2. \hspace{.1in} 	Iterate t times \\
3. \hspace{.20in} 		For each post $i$, sample $z_i,f_i$ \\
4. \hspace{.1in} 	Update joint counts \\
5. \hspace{.1in} 	Iterate until no new post \\
6. \hspace{.20in} 		Read next N posts \\
7. \hspace{.30in} 		For child thread j=1 to K \\
8. \hspace{.30in}	 		Send posts j(N/K) + (j+1)N/K, joint counts \\
9. \hspace{.30in}	 		Receive labels $\{z_i,f_i\}$ for N/K posts\\
10. \hspace{.10in} 	Wait until child threads complete \\
11.\hspace{.10in} 	Iterate t times \\
12.\hspace{.20in} 		For each post with new label $z_i$, sample $z_i,f_i$ \\
13.\hspace{.10in} 	Update joint counts \\
\\
Child Thread \\
1. \hspace{.1in} 	Sleep until invoked by Master Thread \\
2. \hspace{.1in}	Receive N/K posts, joint counts \\
3. \hspace{.1in}	Iterate t times \\
4. \hspace{.20in}		For each post $i$, sample $z_i,f_i$ \\
5. \hspace{.1in}	Return N/K labels $\{z_i,f_i\}$ to Master Thread \\
\hline
\end{tabular}
\end{center}
}
\end{table}

After an initial batch phase (master thread: steps 1-4), the algorithm iterates over three phases: data access (master thread: step 6), computation (child threads: steps 2-5) and synchronization (master thread: steps 8-9, 11-13). 
The initial batch phase is necessary to prevent creation of too many new topics at the beginning by different child threads.
Note that the computation phase happens in parallel across the $K$ child threads.
Each child thread creates multiple new topics, whose counts are maintained locally.
These counts are passed back to the master thread, along with other counts at the end of the computation phase.
After receiving back labels from all child threads, the master thread re-samples labels for all posts assigned new topics by child threads. 
This helps in the consolidation of new topics, many of which may otherwise be quite similar.
The iterations continue until all posts have been processed.

\section{Experimental Results}\label{exp}
In this section we discuss in detail the experiments that we carried out using the proposed D-MRelCRP model on multiple large real social media datasets. 
We evaluate the following aspects of the model: \\
{\bf (a)} Model goodness: Ability to explain unseen data \\
{\bf (b)} Topics and topic labels: Our inference and learning algorithms assign a topic to each post, and also finds a distribution over words for each topic. We evaluate both aspects qualitatively and quantitatively. \\
{\bf (c)} User personalities and their trends: The major distinctive feature of our model is the influence label associated with each post. Using this label, we are able to estimate the user personalities, or the susceptibility of the user to various influencing factors, and their dynamics. We discuss various insights that we were able to find from personality trends. \\
{\bf (d)} Scalability: One of the main strengths of our inference algorithm is the ability to scale to hundreds of millions of data samples. We evaluate how the running time of our multi-threaded implementation scales with data size. \\
{\bf (e)} Relative importance of factors: The MRelCRP and D-MRelCRP models are able to combine together various influence factors and their dynamics. We analyze the usefulness of the different factors for social media analysis. \\
We would like to point out that no other single model is able to perform such a wide array of tasks in social media analysis.
Wherever possible, we make use of available ground truth or surrogates of it for quantitative evaluation and compare against best available baselines. 
However, as regards our main contribution --- discovering user personality trends --- there does not exist any existing algorithm that can perform this.

{\bf Datasets:} We carried out all our experiments on two different datasets:
(1) \textbf{Twitter}: a collection of 360 million tweets crawled between June and December 2009, and 
(2) \textbf{Facebook}: a collection of about 300,000 posts obtained by extracting feeds from publicly available profiles over a span of three months.

{\bf Default Parameter Settings:} 
The hyper-parameters of our online Gibbs Sampler were initialized as : $\alpha$ = 0.1/K+1 and $\beta$ = 0.1. 

{\bf Baselines: } We compare the performance of our models against the following state of the art models that have been shown to be effective for analyzing microblogs. (a) {\bf Latent Dirichlet Allocation} (LDA) \cite{blei03} (b) {\bf Labeled LDA} \cite{dramage10}, and (c) {\bf Timeline} \cite{ahmed:2010}. 
Labeled LDA is not very generally applicable since it makes use of hashtags assigned to the posts to identify topic labels. While this meta-information is available on Twitter, Facebook does not support it. 
The Timeline model is the closest to our model in that it is a non-parametric topic model that also captures topic dynamics. 

\subsection{Model Goodness}
Goodness of a model is evaluated by how well it is able to fit previously unseen data.
Perplexity is a commonly used to measure generalization ability of topic models \cite{blei03}.
It is defined as
the inverse of the geometric mean per-word likelihood: \\
$
\text{Per}(D_{t}) = exp\{ - \sum_{d=1}^{M} log P(\boldsymbol{w_d}) / \sum_{d=1}^{M} N_d \bigr \}
$,
where $N_d$ is the number of words in the $d^{th}$ post in the held-out test set $D_{t}$, and $log P(\boldsymbol{w_d})$  is its log-likelihood. 
Lower values of perplexity indicate better generalization ability.
%
 
We consider two different datasets for this experiment. 
For each model under consideration, we first train it on Twitter data, and then consider as test set a sample of 8 million tweets from the last one month in our dataset.
Similarly, each model is trained on Facebook data, and evaluated on a sample of 40K posts from the last month's activity. 
Perplexities of various models are recorded in \reftab{tab:cluster+perp}.
Note that unlike our model, LDA requires the number of topics to be specified. We set it to the average number of topics discovered by our model across epochs.
Labeled-LDA cannot be applied for Facebook data, since it requires hashtags. 
It can be seen that D-MRelCRP has the least perplexity in both the cases.
Among baselines, Timeline has the best perplexity.   
This demonstrates that capturing both temporal evolution and relationships is important for explaining future data.

\begin{table} \caption{Perplexity and Clustering Accuracy.} \label{tab:cluster+perp}
\centering
\scriptsize
\begin{tabular}{ | c | c | c | c | c | c | }
  \hline
  & \multicolumn{2}{|c|}{Perplexity}&\multicolumn{3}{|c|}{Clustering Acc. (Twitter)}\\
  \hline
  Model & Twitter & Facebook & nMI & R-Index & F1 \\
  \hline
  DMRelCRP & {\bf 1188.29} & {\bf 1562.34} & 0.93 & 0.88 & 0.86 \\
  Timeline & 1582.86 & 1802.9 & 0.81 & 0.72 & 0.73 \\
  L-LDA & 1982.76 & NA & 1 & 1 & 1\\
  LDA & 2932.06 & 3602.0 & 0.55 & 0.52 & 0.48 \\
  \hline
\end{tabular}
\end{table} 




\subsection{Quality of topics}
The D-MRelCRP model assigns a topic label to each post, indicating its category, and also finds a distribution over vocabulary words for each unique topic label, indicating the semantics of the topic.
Our hypothesis is that by modeling the different influences on the users, D-MRelCRP is able to better identify topics.
To evaluate this, we check topic quality in different ways. 
We directly evaluate the topic labels of posts by comparing against a reasonable gold-standard.
Then, we indirectly evaluate the topic labels of posts by using them as features in two prediction tasks.
Finally, we identify significant topic trends and compare them qualitatively with world knowledge.
We provide more details on these three evaluations next.


\paragraph{Clustering posts using topic labels}
In our proposed models, there is a single topic label associated with each post.
This results in a hard clustering of the posts according to topics.
Therefore, one way to evaluate the topic assignment quality is to evaluate the clustering accuracy.
Gold standard clusters of posts is typically hard to obtain.
As an alternative, in the case of Twitter data, we consider hastags as cluster indicators.
Since it is well known that hashtags are often poor indicators of post clusters, we consider only a few authoritative hasgtags as follows.
We collected $\sim 16K$ posts with hashtags corresponding to highly specific topics, such as \#NIPS2009, \#ICML2009, \#bollywood, \#hollywood, \#www2009 etc. 
We consider this as the test set with gold-standard labels for evaluation.

We use three standard metrics to evaluate clustering accuracy - Normalized mutual information, Rand index and F-measure. 
In \reftab{tab:cluster+perp}, we record the performance of our D-MRelCRP model, and those of the three baselines on the Twitter dataset.
Not surprisingly, labeled LDA correctly identifies the clusters all the time, by virtue of taking hashtags as inputs. 
DMRelCRP comes close, in spite not using knowledge of hashtags at all, and performs better than all other models across all the three evaluation metrics.
Further, on closer inspection, we found that the Labeled LDA clustering is not as good as the numbers indicate, and the two proposed models are often better.
For example, DMRelCRP splits the $\sim$3K posts corresponding to the \#movies hashtag into two topics, and separates out posts originating from India.
Comparison using KL-divergence shows this topic to be very similar to the \#bollywood hashtag.
The \#sports hashtag shows a similar split.
Such fine-grained distinction is not possible for Labeled LDA, or Timeline, which do not capture geographic or other influencing factors.


\paragraph{Prediction Tasks}
Since the cluster gold-standards for posts are unreliable for Twitter, and unavailable for Facebook, we additionally perform the following indirect evaluation of topic assignment to posts.
Topic labels are commonly used as reliable low-dimensional features for learning classifiers \cite{blei03}.
We use the topic labels for posts for two representative prediction tasks in social media with reliable gold-standards: predicting post authorship and predicting commenting activity.


\begin{table} \caption{Prediction Task Accuracies.} \label{pred1}
\centering
\scriptsize
\begin{tabular}{ | c | c | c | c | c | }
  \hline
  & \multicolumn{2}{|c|}{Authorship Prediction}&\multicolumn{2}{|c|}{Commenting Prediction}\\
  \hline
  Model & Twitter & Facebook & Twitter & Facebook \\
  \hline
  DMRelCRP & \textbf{0.793} & \textbf{0.734} & \textbf{0.683} & \textbf{0.648}   \\
  Timeline & 0.718 & 0.669 & 0.582 & 0.579\\
  LDA & 0.521 & 0.432 & 0.429 & 0.482 \\
  L-LDA & 0.647 & NA & 0.542 & NA  \\
 \hline
\end{tabular}
\end{table}


\emph{Predicting Authorship: } 
Given a post $p$ and user $u$, this task is to predict if user $u$ is the author of post $p$.
We construct a Twitter test set having 20M tweets from the last 15 days, and a Facebook  test set having 40K posts from the last one month.
For each user, we create training sets for Twitter and Facebook by including as positives all posts authored by that user, and as negatives a equal-sized random sample from posts authored by other users in the recent past. 
As features, we use the topic label of the post inferred by a specific algorithm, and the time-stamp.
We use k-nearest neighbor classifier (k=5), where we consider minimum distance between the post $p$ posts authored by $u$, with KL-divergence as topic distance and number of in-between days as the time difference. 

\emph{Predicting Commenting Activity: } 
Given a post $p$ by some user $v$, the task is to predict if user $u$ comments on the post.
We similarly construct test and training sets from Twitter and Facebook.
As an additional feature, we consider the number of past interactions between users $u$ and $v$. 
We again use a $k-NN$ classifier (k=5) for prediction.

The accuracies for both prediction tasks for different algorithms are recorded in \reftab{pred1}.
It can be seen that DMRelCRP performs significantly better on both datasets. 
The standard topic model baselines, and also Timeline, do not perform very well on this task. 
This shows the usefulness of topics inferred by considering both relationships and temporal evolution. 
Labeled-LDA performs better than LDA, but in spite of using hash-tags, is significantly outperformed by our proposed  approaches.

In summary, the topics inferred using our model are significantly more useful for prediction tasks involving users and posts compared to state of the art topic models. 

\paragraph{Topic Trending and Major Event Detection}
The inferred topic label for each post, in conjunction with the user label, can be used to identify various topic trends.
From the joint counts $n_{k,u,t}$ of the number of posts by user $u$ at time epoch $t$ on topic $k$, the probability $p_{k,u,t}$ of user $u$ posting on topic $k$ at epoch $t$ can be estimated by normalization.
By subsequent aggregation over subsets of users, popularities of different topics across different user segments can be plotted against epochs. 
When a particular topic $k$ dominates over all others in an epoch, we flag that topics as a {\it major event}, and analyze it using the dominant words in the topic distribution $\phi_k$. 
We were able to identify several break-out events using DMRelCRP topics labels, as we describe below. \\
{\em World-wide Events:}  World wide popularity of a topic $k$ at epoch $t$ is estimated by aggregating $p_{k,u,t}$ over all users $u$. 
The major world-wide events discovered by D-MRelCRP include {\it the demise of Michael Jackson} (Jun 30) (top words: mj, michael, dead, singer, jackson, pop), {\it the Fifa World Cup} (Sep 15-30) (football, soccer, fifa, worldcup) and {\it the launch of Google Wave} (Dec 1-15) (wave, invite, google, launching) \\
{\em Geographic Events:} The popularity of a topic $k$ in a specific geography at epoch $t$ is estimated by aggregating $p_{k,u,t}$ only over all users $u$ in that geography. 
{\it Jeff Goldblum's demise} (Jul 1-15) (death, jeff, actor, goldblum, dies, end, era) was detected as major event in Australia and the UK. 

We were able to verify these world-wide and geographic major events using Google Insights\footnote{http://www.google.com/insights/search/}.
The words from the specific topics appeared in the top searches during these specific intervals, world-wide or in the specific geographies. 
We were similarly able to find major events for small networks of users (e.g. official page for Microsoft on Twitter @MSFTNews and its followed pages) and for important individual users (such as @ICML2009). 
In summary, DMRelCRP enables us to discover interesting topic trends and major events at different levels of granularity.

\subsection{Analysis of Influences}
The distinctive aspect of our model is the label $f_i$ indicating the influencing factor behind the $i^{th}$ post. 
It is difficult to evaluate the accuracy of these inferred factors directly.
Instead, we focus on aggregate analysis that can performed using this label, and the rich insights that we were able to unearth using this.

Using the joint counts $n_{k,u,f,t}$ of topics ($k$), users ($u$) and influence factors ($f$) in each epoch ($t$), we can estimate the probability $p_{k,u,f,t}$ of a specific user $u$ posting on a topic $k$ after being influenced by a factor $f$ in epoch $t$, by normalizing appropriately.
On aggregating over topics $k$, the distribution $p_{u,f,t}$ (corresponding to model parameter $\pi_{u,t}$) over factors $f$ indicates the personality of the user $u$ at epoch $t$.
Plotting these distributions over epochs $t$ shows the personality trend for user $u$.  
Since trends over individual anonymous users are not insightful, below we plot aggregate trends over different interesting user subsets. 
For this, we use heat-maps, where the matrix rows correspond to influence factors, columns to epochs and hotter colors indicate higher probability values.

\begin{figure}[t]
  \centering 
 \includegraphics[scale=0.15]{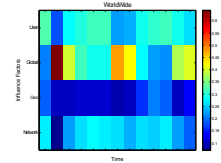}\\ 
  \caption{World-wide personality trends}   
  \label{fig:timeuserhist} 
\end{figure}

{\bf World-wide Personality Trends: }
First, we aggregate $p_{u,f,t}$ over {\it all users} to estimate the world-wide susceptibility of users to specific factors at a specific epoch (15 day period).
This trend is shown in the heat-map of \reffig{fig:timeuserhist} (best viewed in color).
The positions of the hotter colors and the color gradients are of interest.
We can observe that the world-wide factor has the largest variance, followed by personal preferences, while the other trends are largely flat.
Also, we can see that surges in world-wide influence happens mostly as the expense of personal preference.
The largest such surge happens around Jun 30.
This is when the news of Michael Jackson's death broke on Twitter, and we can see that users discarded their personal preferences and posted about this event.
The strength of world-wide influence then subsides gradually, and users return to their personal preferences.
We can see that world-wide influence rises again around Sep 15 and Dec 1, again at the expense of personal preference. 
The most popular topics at these times were {\it FIFA World Cup} and {\it Google Wave}.
In summary, users are usually influenced mostly by their personal preferences and friend network, apart from times of significant world-events. 

\begin{figure}[t]
  \centering 
 \includegraphics[width=0.30\textwidth]{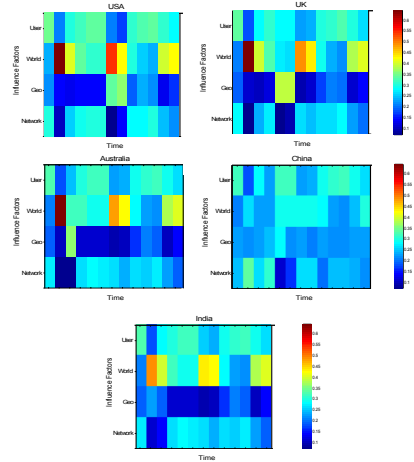}\\ 
  \caption{Personality trends in specific geos}   
  \label{fig:timeuserhist1} 
\end{figure}
{\bf Personality trends in specific geographies: }
Next, we aggregate $p_{u,f,t}$ over {\it users in specific geographies} to estimate susceptibility of users to specific factors in different parts of the world.
Personality trends for 5 different geographies, USA, UK, Australia, China, India, are shown using heat-maps in \reffig{fig:timeuserhist1}.
We can see many interesting patterns.
The personality trends in USA, UK and Australia are largely similar, apart from the geographic influences which are high at different epochs.
For USA, one such high occurs around Sep 15, when {\it US Open} is a dominating topic.
For UK, we can see at high around Aug 15 ({\it Football, Premiere League}) and for Australia around Jul 1 ({\it Jeff Goldblum's demise}).
For India, the relative strengths of world-wide and geographic influences are somewhat weaker. 
For China, the pattern looks different. The strengths of the various influences stay relatively stable, and geographic influence is much stronger than the other 4 cases.

\begin{figure}[t]
  \centering 
 \includegraphics[scale=0.65]{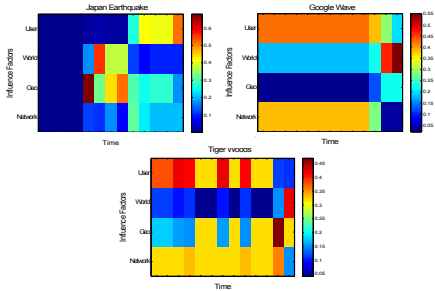}\\ 
  \caption{Character trends of 3 topics}   
  \label{fig:timeuserhist2} 
\end{figure}

{\bf Topic Character Trends: } 
As a final example of the variety of analysis that D-MRelCRP can perform, we look at trends in topic characters. 
By aggregating $p_{k,u,f,t}$ over all users and then using Bayes rule, we can find the posterior distribution $p_{f|k,t}$ over different influence factors for each topic $k$ at epoch $t$. 
By plotting this over epochs, we can see how a topic changes its `character', and moves from a `geographic' topic to a `world-wide' topic, for example.
We illustrate this in \reffig{fig:timeuserhist2} (best viewed in color), using 3 topics.
{\it Japan Earthquake} evolved from a geographic topic to a world-wide topic, {\it Google Wave} from a personal preference topic to a world-wide topic, and {\it Tiger Woods} from a personal preference topic, to a geographic topic, and finally a world-wide topic. 

In summary, DMelCRP enables a wide variety of analysis of influences, leading to many interesting insights, beyond the capability of existing models.

\begin{figure}[t]
  \centering 
 \includegraphics[width=0.30\textwidth]{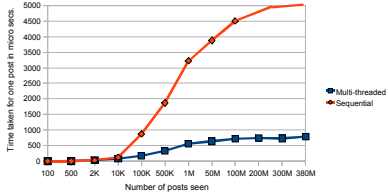}\\ 
  \caption{Scalability of inference algorithms}   
  \label{fig:scalability} 
\end{figure}
\subsection{Other Experiments}
In our experiments, we have employed a Java-based multi-threaded framework over an 8-core, 32 GB RAM machine.
We employed $K=7$ child threads, read $N=35K$ posts in a mini-batch, and used $t=100$ Gibbs iterations per batch.
In \reffig{fig:scalability}, we plot the time taken (in micro-secs) to process one post by the multi-threaded version and a sequential version, after having processed $N$ posts. 
This time increases as the number of living topics increases with $N$.
The two plots clearly demonstrate that superior scalability of our multi-threaded inference algorithm.


\begin{table} \caption{Importance of Model Factors. $R_u$ corresponds to user preferences, $R_w, R_n, R_g$ to world-wide, friend-network and geographic factors, resp.}\label{tab:abl}
\centering
\scriptsize
\begin{tabular}{ | c | c | c | c | c | }
  \hline
  & \multicolumn{2}{|c|}{Perplexity}&\multicolumn{2}{|c|}{Commenting Pred}\\
  \hline
  Model & Twitter & Facebook & Twitter & Facebook \\
  \hline
  DMRelCRP(all) & 1188.29 & 1562.34 & 0.683 & 0.648\\
  MRelCRP(all) & 1345.76 & 1762.01 & 0.602 & 0.538\\
  MRelCRP($R_u,R_n,R_w$) & 1602.86 & 1890.72 & 0.653 & 0.602  \\
  MRelCRP($R_u,R_n$) & 1878.29 & 2245.63 & 0.567 & 0.556 \\
  MRelCRP($R_u,R_w$) & 1802.31 & 2100.01 & 0.512 & 0.54 \\
  RelCRP($R_u$) & 1946.48 & 2189.56 & 0.461 & 0.508 \\
  RelCRP($R_w$) & 2008.59 & 2400.45 & 0.289 & 0.187 \\
  RelCRP($R_n$) & 1958.64 & 2248.90 & 0.478 & 0.423 \\
  RelCRP($R_g$) &  2212.83 & 2890.02 & 0.329 & 0.201 \\
  \hline
\end{tabular}
\end{table}
Finally, in \reftab{tab:abl}, we analyze the contributions of the different aspects to DMRelCRP's final performance. 
We can see that the model improves (both in terms of perplexity and prediction accuracy) through the addition of more relationships and the interplay between them, which is the main strength of the model.


\section{Conclusions}\label{conc}
In this paper, we have made a first attempt at studying the important problem of analyzing user influences in generation of social media data.
We have proposed a new non-parametric model called Dynamic Multi-Relational CRP that incorporates the aggregated influence of multiple relationships into the data generation process as well as dynamic evolution of model parameters to capture the essence of social network data.
Our multi-threaded online inference algorithm allowed us to analyze a collection of 360 million tweets.
Through extensive evaluations, we demonstrated that the topic trends discovered by our model are superior to those from state-of-the-art baselines.
More importantly, we found insightful patterns of influence on social network users, beyond the capability of existing models.




\scriptsize
\bibliographystyle{abbrv}
\bibliography{sai}

\end{document}